\documentclass[12pt]{article}

\usepackage{amssymb}
\usepackage{color}
\usepackage{epsfig,amssymb,amsfonts,amsmath,graphicx,dsfont,cite,xfrac}
\usepackage{authblk}
\usepackage{braket,bbm}
\definecolor{mygray}{gray}{0.5}
\usepackage{subfigure}

\usepackage{cite}
\usepackage[colorlinks=true,linkcolor=blue,citecolor=red]{hyperref}

\newcommand{\be}{\begin{equation}}
\newcommand{\ee}{\end{equation}}
\newcommand{\bea}{\begin{eqnarray}}
\newcommand{\eea}{\end{eqnarray}}

\title{Interaction-Free Measurements of Optical Semitransparent Objects}

\author[]{Zurika Blanco-Garcia}
\author[]{Oscar Rosas-Ortiz}

\affil[]{\footnotesize Physics Department, Cinvestav, AP 14-740, 07000
M\'exico DF, Mexico}

\date{}
\begin{document}

\maketitle

\begin{abstract}
We substitute the fully absorbing obstacle in the Elitzur-Vaidman experiment by a semitransparent object and show that the probabilities of detection can be manipulated in dependence of the transparency of such an object. Then, we connect our results with the delayed choice experiment proposed by Wheeler. It is found that the transparency of the obstacle determines either a particle-like or a wave-like behaviour of a test photon.
\end{abstract}

\section{Introduction}

In quantum mechanics the pure states describe the dynamics of single particles and the measuring devices are macroscopic apparatuses that obey the rules of the classical theory \cite{von32}. One of the most controversial elements of the theory is the {\em axiom of the wave-packet reduction} suggested by the Born's probabilistic interpretation (see e.g. \cite{Mie09} and references quoted therein for details). Indeed, before a measurement the particle is in a superposition of all the possible states associated with a given observable $A$. After measuring $A$, the state $\vert \psi \rangle$ of the particle is not arbitrary anymore, it has been  {\em reduced} to the state $\vert \alpha \rangle$ that corresponds to the measured quantity $\alpha$. Before such a process we have no means to know the specific value of $\alpha$, so the act of measurement involves an unpredictable perturbation of the particle state. To be more precise, let us follow \cite{Mie09} and consider a light beam $\chi_a$ polarized in the direction $\vec e_a$. If the beam impinges on a polarization filter  $A^{\bot}$ that selects polarization $\vec e_a^{\bot}$ perpendicular to $\vec e_a$, no photon will cross $A^{\bot}$. The situation changes if the beam is first transmitted trough a filter $B$ selecting the polarization $\vec e_b$ ($\vec e_a \cdot \vec e_b = \cos \theta$) because, depending on $\theta$, a part of the photons emerging from $B$ will cross the filter $A^{\bot}$. Clearly, the measuring device $B$ not only selects but also affects the state of the photons. 

\begin{figure}[htp]
\centering
\includegraphics[width=0.4\textwidth]{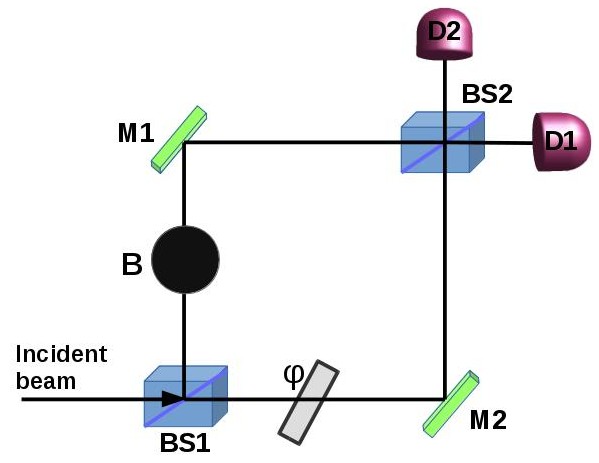}
\caption{\footnotesize The Elitzur-Vaidman experiment for interaction-free measurement. An absorbing object $B$ in one of the arms of the Mach-Zehnder interferometer is `seen in the dark' as a consequence of the quantum  correlations between the wave-packets in the interferometer circuit.}
\label{MZ}
\end{figure}

An intriguing question is if the superposed quantum states are attainable to microscopic systems only. The consequences of assuming that mesoscopic and macroscopic systems can be described by wave-functions were first indicated by Schr\"odinger and gave rise to the concept of quantum entanglement \cite{Sch35}. However, is it possible to identify entanglement in a system as large and complex as the one composed by a living cat and a decaying atom? Although this last is still an open problem, objects as large and classical as the fullerene molecules, the `soccer-ball-shaped carbon cages $C_{60}$', have the ability of forming interference patterns \cite{Arn99,Nai02,Nai03}. On the other hand, the unavoidable influence of the measuring device on the state of a given particle has been harnessed in surprising form by Elitzur and Vaidman  \cite{EV}. They designed an experiment in which a single photon is sent towards a Mach-Zehnder interferometer that includes a fully absorbing obstacle $B$ in one of the arms (see Figure~\ref{MZ}). The quantum state of the photon is divided by a beam splitter ($BS_1$) into two new quantum states represented by wave-packets propagating along the separated space trajectories defined by the arms of the interferometer. The obstacle represents an apparatus that measures the presence of the photon so that this last is either absorbed in the obstructed path or its wave-packet collapses to the free path. In the second case, the photon will be registered by the detector $D_2$ with probability $1/4$. This last result is remarkable because in absence of the obstacle the superpositions are fully destructive in the position of $D_2$, so that the probability of registering the photon in $D_2$ is zero. Thus, the presence of the obstacle is revealed if $D_2$ is activated, no matter that the obstacle was never illuminated by the single photon in the circuit. A phenomenon called `quantum seeing in the dark' \cite{Kwi96}.

In this contribution we explore the consequences of substituting the fully absorbing obstacle $B$ by a semitransparent one in the experimental arrangement of Elitzur and Vaidman. Although some research has been already developed in this trend, see e.g. \cite{Azuma,Jan99,Tho14}, we are interested in the connection between the Elitzur-Vaidman approach and the `delayed choice' experiment proposed by Wheeler \cite{Whe84}. Such an experiment includes a conventional Mach-Zender array in which the observer has the choice of removing the second beam splitter to perform a which-path measurement (i.e., a particle-like experiment). The most interesting part of the Wheeler's proposal is that the removing of $BS_2$ can be done once the photon is already inside the interferometer. In this form, the photon faces either an interference or a which-path experimental setup just when it cannot, in principle, `adjust' its behaviour (wave-like or particle-like) consistently. Although such arrangement was originated as a {\em gedanken} experiment, its quantum predictions have been verified in actual laboratories by using diverse systems, see recent results in  e.g. \cite{Peruzzo,Kaiser}. 

Next, after a brief survey of the Mach-Zehnder interferometer generalities, we recover the Elitzur-Vaidman results by using the Hubbard representation of the operators associated with the devices in  the optical bench. Then we substitute the full absorbing obstacle of Elitzur and Vaidman by a semitransparent one and show that the probabilities of registering the photon in one of the two detectors can be manipulated accordingly. Finally, we show that these results are in connection with the delayed choice experiment of Wheeler.

\section{Quantum states of light in a Mach-Zehnder interferometer}

The spatial states of the photons in the interferometer are elements of the two-dimensional vector space that is generated by the orthonormal basis $\vert r_H \rangle$, $\vert r_V \rangle$, where the sub-label $H$ ($V$) stands for the horizontal (vertical) orientation defined in Figure~\ref{MZ}. The occupation states of the photons are defined by the Fock vectors $\vert 0 \rangle$ and $\vert 1 \rangle$, meaning zero photons and a single photon respectively. Therefore, the quantum states in the circuit of the interferometer are elements of the four-dimensional vector space
\be
{\cal H} = \mbox{Span} \{ \ket{r_H}\otimes\ket{0}, \ket{r_H}\otimes\ket{1}, \ket{r_V}\otimes \ket{0},  \ket{r_V}\otimes \ket{1} \},
\ee
where $A \otimes B$ is the direct product between $A$ and $B$. The mirror $M$, beam splitter $BS$, and phase shifter $\varphi$ indicated in Figure~\ref{MZ} are two-channel operators expressed in Hubbard representation:
\be
M=i(\ket{r_H}\bra{r_V}+\ket{r_V}\bra{r_H}), \hspace{0.5cm} BS=\frac{1}{\sqrt{2}}(\mathbb I_s+M), \hspace{0.5cm}
\varphi= e^{i \phi}\ket{r_H}\bra{r_H}+\ket{r_V}\bra{r_V}.
\ee
Here $\mathbb I_s$ is the identity operator in the vector space ${\cal H}_s = \mbox{Span} \{ \vert r_H \rangle, \vert r_V \rangle \}$, and $\phi \in  [0,2\pi)$. The dyads $X^{k,j} = \vert r_k \rangle \langle r_j \vert$, $r,j, =H,V$, are square matrices of order 2 that have  entry 1 in position $(k,j)$ and zero in all other entries (see \cite{Enr13} for details). These operators are promoted to act in the space ${\cal H}$ as follows
\be
M \leftrightarrow M \otimes \mathbbm{I}_f,\qquad BS \leftrightarrow BS \otimes \mathbbm{I}_f, \qquad \varphi \leftrightarrow  \varphi \otimes \mathbbm{I}_f,
\ee 
with $\mathbb I_f$ the identity operator in the vector space ${\cal H}_f =\mbox{Span} \{ \vert 0 \rangle, \vert 1 \rangle \} \subset \mbox{Span} \{ \vert n \rangle\}_{n\geq 0}$.

\vskip1ex
In the conventional Mach-Zehnder interferometer the entire circuit corresponds to the action of the operator $A=(BS_2) (M \varphi) (BS_1)$ on a given initial state $\vert \psi_{in} \rangle$ which, without loss of generality, will be taken as $\vert \psi_{in} \rangle = \ket{r_H} \otimes \ket{1} \equiv \ket{r_H, 1}$. That is, only one photon (in the horizontal channel) is sent towards the interferometer. After some calculations one arrives at the state
\be
\vert \psi \rangle = A \vert \psi_{in} \rangle= \tfrac{i (e^{i \phi}-1)}{2} \ket{r_V,1}- \tfrac{(e^{i \phi}+1)}{2} \ket{r_H,1}.
\ee
Therefore, the photon will be registered in either $D_1$ (horizontal arm) or $D_2$ (vertical arm) with probability
\be
P_{D_1}= \tfrac{1}{2}(1+ \cos \phi), \qquad P_{D_2}= \tfrac{1}{2}(1-\cos \phi). 
\ee
If the optical path difference of the two arms in the interferometer is zero we can take the phase $\phi=0$ to recover the well known result $P_{D_1}=1$ and $P_{D_2}=0$.

\subsection{The Elitzur-Vaidman model}

To include the presence of a perfect absorbing object $B$ in the first vertical arm we have to identify  the appropriate operator. The boson annihilation operator $a \ket{n}=\sqrt{n}\ket{n-1}$, $n\geq 0$, is useful in this matter. After promoting it as $a \leftrightarrow X^{V,V} \otimes a + X^{H,H} \otimes \mathbb I_f$, one gets
\begin{equation}
\begin{aligned}
\ket{\psi_{in}} &\xrightarrow{\,\,\, BS_1\,\,\, } \tfrac{1}{\sqrt{2}} (\ket{r_H,1}+i\ket{r_V,1}) \xrightarrow{\,\, \, \varphi \, \,\,} \tfrac{1}{\sqrt{2}} (e^{i \phi} \ket{r_H,1}+i\ket{r_V,1})\\[1ex]  
 &\quad \xrightarrow{\,\,\, B \,\,\, } \tfrac{1}{\sqrt{2}} (e^{i \phi} \ket{r_H,1}+i\ket{r_V,0}) \xrightarrow{\,\,\, M \,\,\, } \tfrac{i}{\sqrt{2}} (e^{i \phi} \ket{r_V,1}+i\ket{r_H,0})\\[1ex] 
 &\quad \quad \xrightarrow{\,\,\, BS_2 \,\,\, } \tfrac{1}{2} (i e^{i \phi} \ket{r_V, 1}-e^{i \phi} \ket{r_H, 1}- \ket{r_H, 0}-i \ket{r_V, 0})= \ket{\psi_{EV}}.
\end{aligned}
\label{ecs}
\end{equation}
So that we recover the Elitzur-Vaidman prediction $P_{D_1}= P_{D_2}=\tfrac14$, with probability $1/2$ of finding that the photon is absorbed by the obstacle $B$.

\section{Seeing in the dark a semi-transparent object}

\begin{figure}[htp]
\centering
\subfigure[\footnotesize{$\Delta=0$}]{\includegraphics[width=0.35\textwidth]{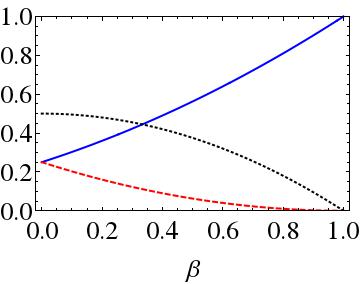}}\hspace{5ex}
\subfigure[\footnotesize{$\Delta=\pi/2$}]{\includegraphics[width=0.35\textwidth]{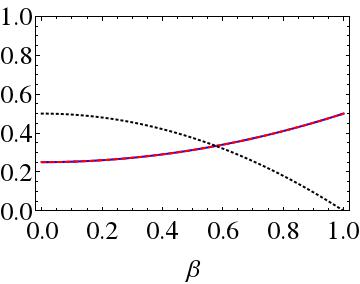}}

\caption{\footnotesize Probability of registering a test photon in $D_1$ (blue, continuous) and $D_2$ (red, dashed) as a consequence of adding an obstacle of transmission coefficient $\Lambda_t = \beta e^{i\theta}$ in the first vertical arm of a Mach-Zehnder interferometer. In (a) and (b) the optical path is such that $\theta=\phi$  and $\theta = \phi + \pi/2$ respectively.  Notice that $P_{D_1}= P_{D_2}$ in (b). The dotted black curve has been included as a reference and corresponds to the probability of absorption $\alpha^2/2$.}
\label{Graf}
\end{figure}

Let us assume that the obstacle $B$ is charaterized by its capability of transmitting or absorbing photons.  If $\Lambda_t$ and $\Lambda_a$ are the related (complex) transmission and absorption coefficients we can represent them in polar form $\Lambda_t = \beta e^{i\theta}$ and $\Lambda_a = \alpha e^{i\gamma}$, with $\beta$ and $\alpha$ nonnegative numbers such that $\alpha^2 + \beta^2 =1$, and the phases $\theta$ and $\gamma$ defining the related principal branches. Therefore, the obstacle will be fully absorbent for $\beta=0$ and transparent for $\beta =1$. Its presence in the interferometer can be associated to the action of the operator $X^{V,V} \otimes (\Lambda_a a + \Lambda_t \mathbb I_f) + X^{H,H} \otimes \mathbb I_f$ on the vector states representing the photon in the circuit. The operation is similar to the one in (\ref{ecs}) and gives
\be
\vert \psi_{abs} \rangle = \tfrac{i(e^{i\phi}-\Lambda_t)}{2} \ket{r_V, 1} - \tfrac{(e^{i \phi} + \Lambda_t)}{2}\ket{r_H,1} - \tfrac{\Lambda_a}{2}  (\ket{r_H, 0} +  \ket{r_V,0}).
\ee
Besides the probability $\alpha^2/2$ of absorption we have
\be
P_{D_1}= \tfrac{1}{4} (1 + \beta^2 + 2 \beta \cos \Delta), \quad P_{D_2} = \tfrac{1}{4} (1 + \beta^2 - 2 \beta \cos \Delta), \quad \Delta = \theta- \phi.
\label{we}
\ee
Clearly, depending on $\beta$ and $\Delta$, we can manipulate the probabilities in order to increase the registering of our test photon in $D_2$ over the possibility of destroying it by the action of the obstacle $B$. As we can appreciate in Figure~\ref{Graf}(a), there is no way to get success if $\Delta =0$. That is, if the optical path in the interferometer arms is such that $\theta=\phi$, the major probability will be associated with either destroying the photon (black dotted curve in the figure) or registering it in $D_1$ (blue continuous curve in the figure), this last giving no information about the presence of $B$ as discussed above. However, if $\theta = \phi + \pi/2$ we have success for $\beta > 0.57$ because $P_{D_1}= P_{D_2} > \alpha^2/2$, see Figure~\ref{Graf}(b).

In general, for a fully absorbing obstacle ($\beta=0$) we recover the Elitzur-Vaidman results, no matter the value of $\Delta$. On the other hand, for complete transparency $\beta=1$ and  $\Delta = 2n\pi$, $n=0,1,\ldots$, only the detector $D_1$ is activated. Therefore we cannot, in principle, distinguish if the obstacle is present or not. However, in this case there is still an arbitrary value of the phase $\theta$ which could be different from zero. Such a phase is inherited to the final state and can be calculated (in principle) by measuring the delay time of the test photon with respect to a similar wave-packet propagating freely and traveling the same optical distance. Our point is that even if the obstacle is transparent we would know about its presence by measuring time-delays in the photons arriving at the zone of $D_1$. If by chance, besides $\beta=1$ we have $\theta =0$, then the obstacle is practically invisible whenever the photon is registered by $D_1$.

\subsection{Approaching the Wheeler gedanken-experiment}
\label{startsample}

\begin{figure}[htp]
\centering
\subfigure[$P_{D_1} (\beta, \Delta)$]{\includegraphics[width=0.45\textwidth]{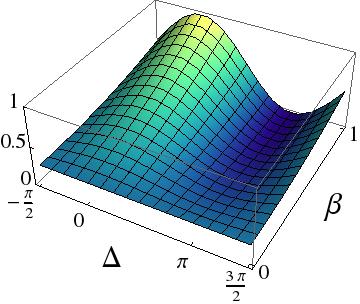}} \hspace{3ex}
\subfigure[$P_{D_2}  (\beta, \Delta)$]{\includegraphics[width=0.45\textwidth]{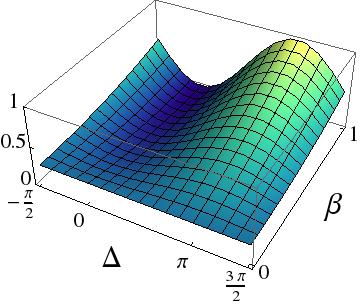}}
\caption{\footnotesize The probabilities (\ref{we}) as a function of $\Delta$ and $\beta$. For $\beta=0$ the photon exhibits particle-like behaviour while $\beta=1$ is associated to wave-like behaviour. Then, depending on $\beta$, the results correspond either to a which-path or to an interferometer experiment.
}
\label{PW}
\end{figure}

Figure~\ref{PW} includes a graphic of the probabilities (\ref{we}) as a function of $\beta$ and $\Delta$. For $\beta=0$ (i.e., a fully absorbing obstacle) the probabilities of registering the photon by $D_1$ and $D_2$ are the same and equal to $1/4$. This last because the photon wave-packet  collapses to the free path in the interferometer and the beam splitter $BS_2$ decouples it in a pair of coherent packets.  Thus, the photon exhibits a particle-like behaviour because the result corresponds to a which-path experiment. A different situation arises if the obstacle is transparent ($\beta=1$) because, depending on the optical path of the interferometer (i.e., depending on $\phi$),  the interference of the outgoing packets is such that the major probability of registering the photon alternates between $D_2$ and $D_1$. In this case the photon exhibits a wave-like behaviour as the result corresponds to an interference experiment. For other values of $\beta$ and $\Delta$ there is not a clear distinction between the particle-like and wave-like behaviour of the photon. Our results are in qualitative agreement with the recent (experimental and theoretical) results reported in e.g. \cite{Peruzzo,Kaiser}, where the authors use a controlled-Hadamard operation to simulate a movable beam splitter $BS_2$.

\section{Conclusions}

We have shown that the Elitzur-Vaidman model of interaction-free measurements can be connected with the delayed choice experiment suggested by Wheeler. The main point is the substitution of the fully absorbing obstacle of Elitzur and Vaidman by a semitransparent one. Actual implementations of experimental arrays oriented to verify in the laboratory our quantum predictions are in progress (for details see \cite{tesisZB}).

\subsection*{Acknowledgments}
The comments and suggestions by the anonymous referees are acknowledged. ZBG gratefully acknowledges the funding received through a CONACyT Scholarship.


\end{document}